\documentclass[aps,prb,twocolumn,showpacs,groupedaddress,citeautoscript]{revtex4}  
\usepackage{graphicx}  
\usepackage{dcolumn}   
\usepackage{bm}        
\usepackage{amssymb}   
\usepackage{graphicx}
\usepackage{color}
\usepackage{bm}
\usepackage{amssymb}
\usepackage{amsmath}
\usepackage{subfigure}
\usepackage{tikz}
\usetikzlibrary{shapes,arrows}
\usepackage{amsmath}
\usepackage{appendix}
\usepackage{multirow}
\usepackage{booktabs}

\def\cFrac#1#2{%
\begin{array}{@{}c@{}}\multicolumn{1}{c|}{#1}\\%
\hline\multicolumn{1}{|c}{#2}\end{array}}

\begin{document}



\title{Renormalization of two-body interactions due to higher-body interactions of lattice bosons}
\author{Vipin Kerala Varma} \affiliation{Bethe Center for Theoretical Physics, Universit\"{a}t Bonn, Germany} \affiliation{The Abdus Salam International Centre for Theoretical Physics, Trieste, Italy}
\author{Hartmut Monien} \affiliation{Bethe Center for Theoretical Physics, Universit\"{a}t Bonn, Germany}

%
\date{\today}

\vspace*{-1cm}
\begin{abstract}
We calculate thermodynamic properties of soft-core lattice bosons with on-site $n$-body interactions using up to twelfth and tenth order 
strong coupling expansion in one and two dimensional cubic lattices at zero temperature. Using linked cluster techniques, we show that it is 
possible to exactly renormalize the two-body interactions for quasiparticle excitations and ground-state energy by resumming the three and 
four body terms in the system, which suggests that all higher-body on-site interactions may be exactly and perturbatively resummed into the 
two-body terms for similar system observables. The renormalization procedure that we develop is applicable to a broad range of systems 
analyzable by linked cluster expansions, ranging from perturbative quantum chromodynamics to spin models, giving either an exact or approximate 
resummation depending on the specific system and properties. 
Universality at various three-body interaction strengths for the two dimensional boson Hubbard model is checked numerically.
\end{abstract}

\maketitle
\section{Introduction}
The first calculation of the effect of three-body interactions in lattice bosons \cite{Murphy} for liquid 
He$^{4}$ and solid He revealed its negligible effect on its ground state energy. However, it was recently 
suggested \cite{Buechler} that, firstly, three-body interactions in cold polar molecules could be naturally 
modelled by Hubbard Hamiltonians with nearest-neighbour three-body terms and, secondly, there might arise 
new exotic phases in experimental setups of such degenerate quantum molecular gases. Shortly thereafter, a 
decoupling mean-field (MF) approximation was used to investigate the critical properties of a boson Hubbard 
model with \textit{on-site} three-body interactions \cite{Zhang}. That such on-site terms could effectively 
arise in two-body collisions of atoms confined to optical lattices was only subsequently justified \cite{Johnson}. 
Indeed such a multi-body interaction Hamiltonian, of the form Eq. \eqref{hamiltonian} that we study, was observed through
quantum phase revivals in a system of ultracold $^{87}\textrm{Rb}$ atoms with virtual transitions from the lowest vibrational state 
to higher energy bands \cite{Will}.\\
We employ a strong coupling expansion $-$ which suffers from no finite size effects as it operates directly 
in the thermodynamic limit $-$ of the on-site three-body interacting boson Hubbard model, in 
addition to checking the universality hypothesis at $XY$ critical points for various three-body strengths. 
A procedure for incorporating higher body interactions by renormalizing the two-body problem will also be 
described herein.\\ 
Consider a system of two-body and higher-body interacting bosons on the one dimensional chain and two dimensional 
square lattices described by the Hamiltonian
\begin{eqnarray}
 H &=& -t\sum_{<i,j>}(b_i^{\dagger}b_j^{\phantom{\dagger}} + \textrm{h.c}) + \frac{U}{2}\sum_{i}\hat{n_i}(\hat{n_i}
 - 1) \nonumber \\
& &+ \sum_{k=3}\frac{U_{k}}{k!}\sum_{i}\prod_{l=0}^{k-1}(\hat{n_i} - l)- \mu\sum_{i}\hat{n_i}.
\label{hamiltonian}
\end{eqnarray}
where the $b_i^{\dagger}$ and $b^{\phantom{\dagger}}_i$ are bosonic creation and annihilation operators, 
$\hat{n_i}= b_i^{\dagger}b_i^{\phantom{\dagger}}$ is the number operator, the hopping-terms $t$ are between 
nearest neighbors, and the system consists of a single species of soft-core bosons. The local energy term $U$ 
contributes to a repulsive on-site interaction between bosons, $U_k > 0$ is the strength of $k$-body on-site 
interaction terms and $\mu$ is the chemical potential. The onsite term $U$ will be the energy scale of choice in 
this letter. \\
The Hamiltonian in Eq.~(\ref{hamiltonian}), when represented in the form $\mathcal{H} = \mathcal{H}_0 - \lambda 
\mathcal{H}_{1}$ with $\mathcal{H}_{1}$ being the hopping terms of strength $\lambda = t/U$ and $\mathcal{H}_{0}$ 
being the rest of Eq.~(\ref{hamiltonian}), is amenable to linked cluster expansions \cite{Singh, Elstner} in the 
parameter $\lambda$. Evaluation of physical properties e.g. energy are performed via Rayleigh-Schr\"{o}dinger 
perturbation theory \cite{Baym} and the linked cluster expansion. Excited states can be obtained using a similar 
procedure through Gelfand's similarity transformation \cite{Singh}.\\
\section{Critical properties}
In this section we focus on the critical properties of the transition between the ground state Mott insulator and the superfluid 
phase in the Bose-Hubbard model. The quantum phase transition at the tip of the insulating lobe will be a special point of concern 
because the system's universality with the XY model may be investigated at this multicritical point \cite{Fisher, Freericks}.
\subsection{One dimensional chain lattice}
\begin{figure}[th!]
\includegraphics[scale=0.32]{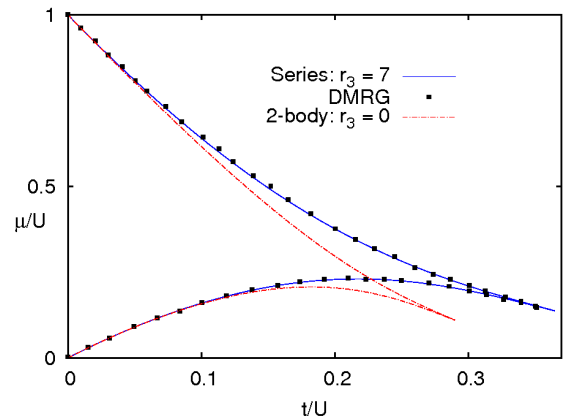}
\caption{\label{1DPhase} First Mott lobe in the ground state phase diagram of the one dimensional boson Hubbard 
model with various $r_3 \equiv \frac{U_3}{U}$ ratios; the last Pad\'{e} approximant [6/6] or [5/5] of the twelfth 
order gap series for $r_3 = 7$ is compared with a DMRG solution \cite{Valencia}. [m/n] denotes an $m^{\textrm{th}}$
order numerator and an $n^{\textrm{th}}$ order denominator. The Kosterlitz-Thouless point shifts upwards and 
rightwards as $r_3$ is increased.}
\end{figure}

Consider first the $\rho = 1$ Mott insulating lobe in the one dimensional chain. For a twelfth order 
bond-expansion, there are 13 distinct topological graphs (clusters) that can be embedded on the infinite chain: 
approximately 2.5 million states contribute to the full Hilbert space with a maximum of 13 states in the lowest 
degenerate manifold. From MF calculations \cite{Zhang, Shou}, it was predicted that the first Mott lobe should 
not change in structure; this was later systematically corrected by density matrix renormalization group (DMRG) 
calculations \cite{Valencia, MSingh} and exact diagonalization \cite{Sowinski}. Using Gelfand's similarity transformations to construct the particle and 
hole excitations, we identify the disappearance of the excitation gap as defining the second-order transition 
contours of the lobe: we have thus evaluated the series for the gap up to twelfth order; moreover we emphasize that because we work in the 
thermodynamic limit, there are no finite 
size effects in the sense that each of the coefficients in the series are exact to any given perturbative order. To illustrate, 
for $r_3 \equiv \frac{U_3}{U} = 1$, the first eight terms in the Mott gap, $\delta_{1}(\lambda, {\bf k} = 0)$, 
are given by
\begin{widetext}
\begin{eqnarray}
 \delta_1(\lambda, {\bf k} = 0) = 1 - 6\lambda + 6\lambda^2 + \frac{20}{3}\lambda^3 - \frac{46}{27}\lambda^4 + 
 \frac{30751}{243}\lambda^5 - \frac{185083}{324}\lambda^6 + \frac{464023295}{157464}\lambda^7 - 
 \frac{68401014192209}{3769688160}\lambda^8,
\label{gap}
\end{eqnarray}
\end{widetext}
where ${\bf k}$ is the lattice momentum. In Fig.~\ref{1DPhase} we show particle and hole contours obtained by 
multiple precision Pad\'{e} approximation of twelfth order series and compared to a previously published \cite{Valencia} DMRG 
solution for $r_3 = 7$. The location of the critical point (Kosterlitz-Thouless transition) shifts upwards and 
rightwards in the phase diagram indicating an increase in the size of the first lobe and a partial restoration of 
particle-hole symmetry as the semi-hardcore condition ($r_3 = \infty$) is reached. We have verified this tendency 
with $r_3 = 0, 1, 7, 100$. For the hardcore condition, the critical $\mu/U$ ($U$ now being the nearest neighbour 
repulsion) equals exactly 1 \cite{Yang}, with the particle-hole symmetry completely restored.\\
We note that in an $n$-th order bond expansion for the linear chain, we need calculate the $n$-th order particle 
and hole contributions only up to graphs with $n-1$ bonds: the effective Hamiltonian for the last cluster may be 
calculated with very little effort because, within the degenerate subspace of this graph, the only contributing 
process will be the one which transfers the excitation from one end of the chain to the other. We find that these matrix 
elements are simply given by the negative of the Schroeder numbers $ S = \{1, 2, 6, 22, 90 \cdots\}$ with the 
generating function \cite{Sloane}
\[
 G(x) = \frac{1 - x - \sqrt{1 - 6x + x^2}}{2x}.
\]
\noindent That is, for a cluster with $n$-bonds the $n^{\textrm{th}}$ order effective Hamiltonian has its non-zero 
elements given by $H(0,n) = -S_{n-1}$, for $l\ge2$. This is independent of $r_3$ and the type of excitation.\\
\subsection{Two dimensional square lattice}
In two dimensions there are 680 topologically distinct clusters contributing at tenth order for a bond-expansion. 
Here the Mott gap closes as $\delta \propto (t_c - t)^{z\nu}$ \cite{Fisher} for $t - t_c \ll 1$, assuming the 
universality of the $XY$ model: $t_c$ is the value of the hopping element at the multicritical point where 
particle-hole symmetry is restored (here $z = 1$), $z$ and $\nu$ are the dynamical and coherence length critical 
exponents respectively. To investigate the effect of three-body terms in 2D, we have calculated tenth order series 
for the $\rho = 1$ Mott gaps for $r_3 = 1, 10, 100$. From these series', $t_{c}$ and $z\nu$ may be extracted by 
proceeding, \textit{mutatis mutandis}, as outlined in previous scaling analysis \cite{Elstner, Varma}: (a) by linearly 
extrapolating the roots of the truncated gap series' from, say, fourth to tenth order and (b) by Pad\'{e} 
approximating the gap series' to mimic the expected $\delta$ behaviour mentioned above. The results of the higher 
approximants ([4/4], 
[4/5], [4/6], [5/4], [5/5]) and linear extrapolation are tabulated in Table \ref{Critical} for four $r_3$ values: 
it must be noted that large $r_3$ values may be attained, as suggested by Johnson {\sl et. al.}, using Feshbach 
resonances and tuning the lattice potential. As can be noted from the table that the change in $t_c$ upon 
increasing $r_3$, and hence the structure of the first lobe, is not as substantial for the square lattice as was 
seen for the one dimensional case. From the $\nu$ values for the four three-body interaction strengths, we see 
that universality does indeed seem to hold at the $XY$ point. The corresponding classical critical coefficient for 
the three dimensional $XY$ model is $\nu = 0.67155 \pm 0.00027$ \cite{Campostrini}.\\
\begin{table}[bh!]
\caption{\label{Critical}Critical points and exponents for the two dimensional square lattice boson Hubbard model 
at various three-body interactions $r_3$ using roots extrapolation (E) and Pad\'{e} approximations (P). See text 
and Refs. 7 and 17 for the exact procedure. At the critical point $t = t_c$, $z = 1$ \cite{Fisher}.}
\begin{ruledtabular}
\begin{tabular}{llll}
$r_3$ & $t_{c}^{\textrm{E}}$($10^{-4}$) & $t_{c}^{\textrm{P}}$($10^{-4}$) & $\nu$($10^{-3}$)\\
\hline
0\footnote{From  Ref. 7} & 597.4 $\pm$ 0.4 & 599 & 690 \\
1 & 603.8 $\pm$ 0.8 & 604.69 $\pm$ 0.06 & 692.3 $\pm$ 0.6\\
10 & 616.7 $\pm$ 0.8 & 617.39 $\pm$ 0.05 & 695.4 $\pm$ 0.4\\
100 & 621.4 $\pm$ 0.8 & 621.98 $\pm$ 0.06 & 696.5 $\pm$ 0.5\\
\end{tabular}
\end{ruledtabular}
\end{table}
\section{Renormalization procedure}
In general, to incorporate a second variable like $U_3$ into a Hamiltonian within linked cluster expansions 
requires a double-expansion: the first in $t/U$, the second in $U_3/U$. For example, the double-expansion of a 
quantity $P$ in perturbing variables $\lambda$ and $r$ to order $M$ and $N$ respectively may be symbolically 
written as
\begin{eqnarray}
P = \sum_i^{M}c_{1i}^{(N)}\lambda ^{i}, \nonumber \\
 c_{1i}^{(N)} = \sum_j^{N}c_{2j}r^j.
\label{doublesum}
\end{eqnarray}

Now $M$ and $N$ are finite integers but can one do better? The prescription we adopt is to resum the second 
series and evaluate $\lim_{N \to \infty}c_{1i}^{(N)}$ for every $i$, keeping $M$ finite, and is implemented 
as follows: we first calculate the series coefficients for a given observable (like in Eq.~\ref{gap}) for a 
finite number of $r_3$ values. And because the coefficients are always rational numbers $-$ by virtue of the 
perturbation theory $-$ it only remains to find a rational function approximation to the obtained coefficients. 
The latter step may be easily implemented with Thiele's algorithm for continued fraction representation 
\cite{Stoer}. \\
\subsection{Thiele's algorithm}
\label{sec: Thiele}
Thiele's algorithm is used to interpolate a given set of support points $(x_i, f_i)$ by a rational function of 
the form
\begin{equation}
\label{eq: RA}
\mathbf{\phi}^{(M,N)}(x) = \frac{P^{(M)}(x)}{Q^{(N)}(x)} = \frac{a_0 + a_1x + \cdots + a_Mx^M}{b_0 + b_1x + \cdots + b_Nx^N},
\end{equation}
for integer coefficients $a_i, b_i$ and some order of the polynomials $M, N$. As with the construction of 
Pad\'{e} approximants, the maximal degree of the numerator and denominator in Thiele's rational function 
approximation are determined by the number of data points available. We closely follow the discussion in 
Ref. 19 in this subsection.\\
Rational expressions are constructed along the main diagonal of the $(M, N)$-plane in Thiele's algorithm. 
The support points $(x_i, f_i)$ are used to generate inverse differences $\phi$ depicted notationally 
in Table \ref{t: ThielesTable}. The inverse differences and the algorithm are defined by the following recursion relation and 
identity \cite{Stoer}
\begin{eqnarray}
\label{eq: CFE}
 \phi(x_i,x_j) &=& \frac{x_i-x_j}{f_i-f_j}, \nonumber \\
 \phi(x_i, \ldots, x_l,x_m,x_n) &=& \frac{x_m-x_n}{\phi(x_i, \ldots ,x_l,x_m) - \phi(x_i, \ldots ,x_l,x_n)}, \nonumber \\
\phi^{(n,n)}(x) &=& f_0 + \cFrac{x-x_0}{\phi(x_0,x_1)} \nonumber \\ &+& \cFrac{x-x_1}{\phi(x_0,x_1,x_2)} + \ldots  \nonumber \\
&+& 
\cfrac{x-x_{2n-1}}{\phi(x_0,x_1, \ldots, 2n)}.
\end{eqnarray}
The last three lines in \eqref{eq: CFE} gives the continued fraction expansion, in Pringsheim's notation, for the Thiele's rational approximation of 
the $2n+1$ data points.\\
\begin{table}[bh!]
\caption{\label{t: ThielesTable}Generic flow of the Thiele's algorithm in construction of inverse differences from the input data set.}
\setlength{\tabcolsep}{0.5em}
\centering
\begin{tabular}{ccl}
 $x_i$&$f_i$&Inverse differences\\ \hline
 $x_0$&$f_0$&\\
 $x_1$&$f_1$&$\phi(x_0,x_1)$\\
 $x_2$&$f_2$&$\phi(x_0, x_2)$\hspace{2em}$\phi(x_0,x_1,x_2)$\\
 $x_3$&$f_3$&$\phi(x_0, x_2)$\hspace{2em}$\phi(x_0,x_1,x_2)$\hspace{2em}$\phi(x_0,x_1,x_2,x_3)$\\
 $\vdots$&$\vdots$&$\vdots$\\
\end{tabular}
\end{table}
\begin{table*}[th!]
\caption{\label{resummed}Resummed series coefficients for the particle and hole contours in the one dimensional 
chain and two dimensional square lattice. Coefficients of lower order that are not listed are independent of $r_3$.}
\setlength{\tabcolsep}{0.5em}
\begin{ruledtabular}
\begin{tabular}{ccc}
 &\multicolumn{2}{c}{Lattice}\\
 Coefficient&\multicolumn{1}{c}{1D}&\multicolumn{1}{c}{2D}\\ \hline
 $c_4^-$&$\cfrac{60-4r_3}{3+r_3}$&$-8\cfrac{231+71r_3}{3+r_3}$\\
 $c_2^+$&$2\cfrac{1+2r_3}{2+r_3}$&$-4\cfrac{7+2r_3}{2+r_3}$\\
 $c_3^+$&$12\cfrac{2+2r_3+r_3^2}{(2+r_3)^2}$&$-24\cfrac{20 + 18r_3 + 3r_3^2}{(2+r_3)^2}$\\
 $c_4^+$&$-2\cfrac{339 + 1631r_3 + 2818r_3^2 + 2088r_3^3 + 676r_3^4 + 80r_3^5}{(2+r_3)^3(3+r_3)(5+4r_3)}$&$-4
 \cfrac{28497+71317r_3+70166r_3^2+33672r_3^3 + 7772r_3^4 + 688r_3^5}{(2+r_3)^3(3+r_3)(5+4r_3)}$\\
\end{tabular}
\end{ruledtabular}
\end{table*}
In the event that one or more of the inverse differences in Table \ref{t: ThielesTable} are equal, then the 
continued fraction expansion must terminate at this column lest the succeeding inverse differences become 
undefined; this abrupt termination usually indicates that the obtained approximation is in fact an exact 
functional representation of the input data.\\
To summarize, the functional dependence of a coefficient at a given order on $r_3$ is to be captured by a 
rational approximation. For example, a single-expansion coefficient $c_{1i}$, for a given $i$, for some 24 
values of $r_3$ from $0 \rightarrow 100$ were evaluated. Thiele's algorithm to find an approximating rational 
function $f_i(r_3) = c_{1i}(r_3)$ would generally require as many steps as there are points (here 24) to 
terminate and find the best fit; however, we find that in each of the evaluated coefficients, the algorithm 
stops exactly after a few steps because the continued fraction expansions stop. This ensures the exactness of 
the obtained $f_i(r_3)$. With this, the $c_1$'s in Eq.~\ref{doublesum} get fully renormalized by the resummed 
$c_2$'s.\\ 
\subsection{Three body interactions}
\begin{figure}[t!]
\includegraphics[scale=0.32]{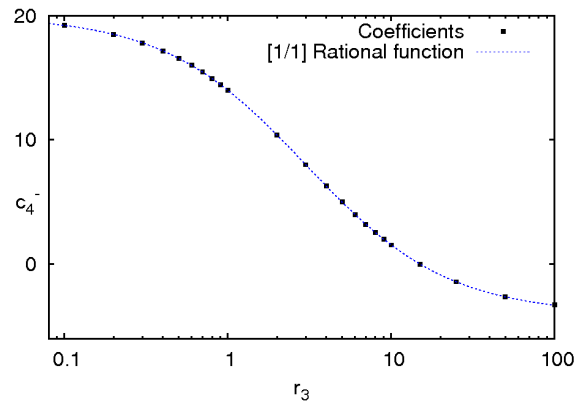}
\caption{\label{c41D} Fourth order coefficient for the hole contour in the one dimensional chain as a function of 
the three-body interacting strength in a log plot. The coefficient at $r_3=0$ passes through the function as well. 
The [1/1] function for $c_4^-$ is $\frac{60 - 4r_3}{3 + r_3}$.}
\end{figure}

The procedure in Sec. \ref{sec: Thiele} can now be applied to renormalizing the series coefficients of the particle and hole contours 
in the two-body interacting one dimensional chain and two dimensional square lattice with respect to the three 
body terms. Let the particle and hole contours, for any $r_3$, be represented as
\begin{equation}
 \pm \mu_{\pm}^{c}(r_3) = \sum_{i=0} c_{i}^{\pm}(r_3)\lambda^{i}.
\end{equation}
The signs ($\pm$) refer to the particle and the hole contours respectively. For illustrating our method, we 
sketch in Fig.~\ref{c41D} the [1/1] rational function approximation to $c_4^{-}(r_3)$ in the one dimensional 
chain obtained from the 24 different values of $r_3$. The same analyses were performed for the particle 
coefficients as well and similar conclusions hold; the resummed coefficients are listed in Table \ref{resummed} 
up to fourth order. For example, in the 1D case, $c_4^-(r_3=1) + c_4^+(r_3=1) = -\frac{46}{27}$, the fourth 
coefficient in Eq.~\ref{gap}. It is worth noting that even with coefficients for particle-hole series only up 
to third order, very reasonable estimates (within 10\% compared to more accurate results \cite{Elstner}) for 
critical properties can be obtained \cite{Freericks}. A similar resummation may be readily obtained, exactly 
and without using the above resummation procedure, for the ground state energy per site $\langle \psi |\mathcal{H}|\psi 
\rangle$ $-$ where $|\psi\rangle$ is the ground 
state wavefunction constructed order by order for the first lobe $-$ in the one dimensional chain for an arbitrary 
$r$ up to fourth order to give 
\begin{equation}
\label{eq: GSE}
 E_0^{1D} = -4\lambda^2 + 12\frac{1+r_3}{3+r_3}\lambda^4.\\
\end{equation} 
It has been checked that the result \eqref{eq: GSE} is also obtained by employing the renormalization procedure described in 
Sec. \ref{sec: Thiele}. \\
We see from Table \ref{resummed} and Eq. \eqref{eq: GSE} that for certain values of attractive interactions i.e. 
$r_3 < 0$ there is a perturbative instability of the $\rho = 1$ Mott phase coming from the divergence of the 
denominators. This might signal the disappearance of the first lobe altogether or the appearance of a 
higher-density and energetically more favourable lobe in that region of phase space: quite naturally, for 
attractive bosons, higher density Mott phases should stabilize the system and one should expand thermodynamic 
variables perturbatively about this more favourable phase. Similar conclusions were in fact reached by recent 
MF and quantum Monte Carlo calculations \cite{Naini}. In the present work, however, the value of the attractive 
three-body strength that leads to an instability at a given perturbative order can be readily read off from the 
resummed coefficients.\\
\subsection{Four body interactions}
Using the above procedure for the Hamiltonian with four body interactions, with $r_4 \equiv \frac{U_4}{U}$, we 
find that the two-body interactions for the ground state energy densities of the linear chain and the square 
lattice may also be perturbatively renormalized by the four-body terms as given by the following:
\begin{widetext}
 \begin{eqnarray}
  E_0^{\textrm{1D}} &=& -4\lambda ^2 + 4\lambda^4 + \frac{272}{9}\lambda^6 + \frac{4(85r_4-7602)}{81(r_4+6)}
  \lambda ^8 - \frac{2(252109r_4^3 + 2870730r_4^2 + 6509628r_4 - 9540936)}{729(r_4+6)^3}\lambda ^{10} + \cdots, 
  \nonumber \\
E_0^{\textrm{2D}} &=& -8\lambda ^2 - 24\frac{7r_4 + 27}{r_4 + 3}\lambda ^4 - 32\frac{514r_4^4 + 6333r_4^3 + 
28167r_4^2 + 53160r_4 + 35217}{(r_4 + 3)^3(2r_4 + 3)}\lambda ^6 + \cdots .
 \end{eqnarray}
\end{widetext}
Therefore it seems very likely that two-body terms in the Bose-Hubbard model, irrespective of dimension, may be 
perturbatively renormalized by \textit{all} higher-body on-site terms for its thermodynamic and excited properties. An interesting 
question is if such resummability might also exist for dynamical properties and for bosonic models with intersite interactions.\\
\section{Summary}
In conclusion, we have presented a way of resumming the effect of a second perturbing variable to infinite order 
thereby effectively renormalizing the series coefficients of the single variable expansions. The procedure is 
quite general and may be applied to renormalize the second interaction term in the Hamiltonian in the series 
expansion representation of any thermodynamic quantity. In the scenario considered, we have found perturbative 
renormalizations of the two-body interactions due to three and four body terms in calculations of ground state 
energies and quasiparticle excitations, for the one and two dimensional Bose-Hubbard model. The applicability of 
the procedure is, of course, not restricted to lattice bosons but can be extended to other systems that are 
treated using series expansion techniques, ranging from spin models to perturbative quantum chromodynamics where 
the analytic continuation of strong coupling expansions is still fraught with problems \cite{Oitmaa}. Additionally,
the universality hypothesis of the two dimensional Bose-Hubbard model has been checked for various three body 
interaction strengths.\\
\section{Acknowledgment}
%
One of us (VKV) thanks the Bonn-Cologne Graduate School for support within the Deutsche Forschunggemeinschaft's Research Funding.
%

\bibliographystyle{unsrt}
\bibliography{Ref6}

\end{document}